\documentclass{article}\usepackage{jkas2}
\runningauthor{RUDNICK}
\runningtitle{Observing Large Scale Fields}
\beginpage{1}
\endpage{5}

\begin{document}

\font\twelvei = cmmi10 scaled\magstep1
       \font\teni = cmmi10 \font\seveni = cmmi7
\font\mbf = cmmib10 scaled\magstep1
       \font\mbfs = cmmib10 \font\mbfss = cmmib10 scaled 833
\font\msybf = cmbsy10 scaled\magstep1
       \font\msybfs = cmbsy10 \font\msybfss = cmbsy10 scaled 833
\textfont1 = \twelvei
       \scriptfont1 = \twelvei \scriptscriptfont1 = \teni
       \def\mit{\fam1 }
\textfont9 = \mbf
       \scriptfont9 = \mbfs \scriptscriptfont9 = \mbfss
       \def\bmit{\fam9 }
\textfont10 = \msybf
       \scriptfont10 = \msybfs \scriptscriptfont10 = \msybfss
       \def\bmsy{\fam10 }

\def\etal{{\it et al.~}}
\def\eg{{\it e.g.,~}}
\def\ie{{\it i.e.,~}}
\def\lsim{\raise0.3ex\hbox{$<$}\kern-0.75em{\lower0.65ex\hbox{$\sim$}}}
\def\gsim{\raise0.3ex\hbox{$>$}\kern-0.75em{\lower0.65ex\hbox{$\sim$}}}

\title{Observing Magnetic Fields on Large Scales}

\author{Lawrence Rudnick}
\address{Department of Astronomy,
University of Minnesota, Minneapolis, MN, USA\\
{\it E-mail: larry@umn.edu}}

\address{\normalsize{\it (Received October 31, 2004; Accepted December 1,2004)}}

\abstract{Observations of magnetic fields on scales
up to  several Mpc are important for understanding cluster and large-scale structure
evolution. Our current census of such structures is heavily
biased  -- towards fields of several $\mu$G,
towards fields in deep potential wells, and towards high inferred
field strengths in cooling flow and other clusters from improper
analysis of rotation measure data.  After reviewing these biases,
I show some recent results on two relics that are powered in very
different ways.  I describe new investigations that are now uncovering weak
diffuse fields in the outskirts of clusters and other low density
environments, and the good prospects for further progress.}


\maketitle

\section{Introduction}

Studies of diffuse extragalactic magnetic structures -- those not
 clearly associated with parent AGNs -- have reached a level
 of sufficient  maturity that it is time to review what we've seen
 and what we might have missed.
 I will start by examining the observations in the phase space 
of magnetic field strength and angular size [B,$\theta$].  After using this to
discuss some of the inherent observational biases in magnetic field studies, I
will highlight three different types of diffuse structures of
current interest - cluster-wide fields probed through rotation
measures, energization of cluster ``relic" sources, and
diffuse sources seen in weak potential wells.

\section{[B,$\theta$] phase space}
 Figure 1 introduces
[B,$\theta$] space where we can 
look at a) known objects, b) benchmark fields that would
be pressure matched with cosmological structures, and c) selection effects and
other biases. The exact
placement of features on this diagram is not important 
at this stage --  increasing angular size generally
reflects physically larger structures. However, since the
selection effects depend only weakly on redshift,
[B,$\theta$] space is a  useful place to start our investigation.

Beginning with the observations, we note that most information
about magnetic field strengths comes from minimum energy estimates
(B$_{min}$)
-- {\it i.e.,} the field strength that minimizes the total energy
in the relativistic plasma under the constraint of the observed
synchrotron luminosity. This closely approximates the field
strength that yields an equipartition in energy between fields and
relativistic particles. Readers are cautioned to distinguish such
field-particle equipartition from other uses of the term, such as
the equipartition of pressure between relativistic and thermal
plasmas.

At the small, high field (pressure) end, we find the hot spots of powerful
radio galaxies and quasars --these are transient features associated
with jet-driven shocks, pressure balanced on average by the ram
pressure of their advance into the local thermal medium.  In
clusters of galaxies, we find larger scale tails and bridges of
emission associated with AGN outflows.  These diffuse structures
presumably move only slowly through their local medium and are
therefore approximately in static pressure balance. However, the
minimum pressure estimates often fall below those of the surrounding
medium (e.g., Morganti et al. 1998;  Worrall \& Birkinshaw 2000), so
additional sources of pressure are needed.  At even lower
minimum energy fields, we find the radio ``relics'', discussed
further below and in these proceedings in more detail by G.
Giovannini.  Cluster-wide halos, as discussed here by L. Ferretti,
have minimum energy or inverse-Compton estimated field strengths
in the range 0.1 - 1$~\mu$G.  I have not included in the [B,$\theta$]
diagram  the 1-35~$\mu$G  rotation
measure estimates of cluster fields
(Carilli \& Taylor 2002)
since I do not believe they are supported by the observational
evidence.
\begin {figure*}[t]
\vskip 0cm \centerline{\epsfysize=10.cm\epsfbox{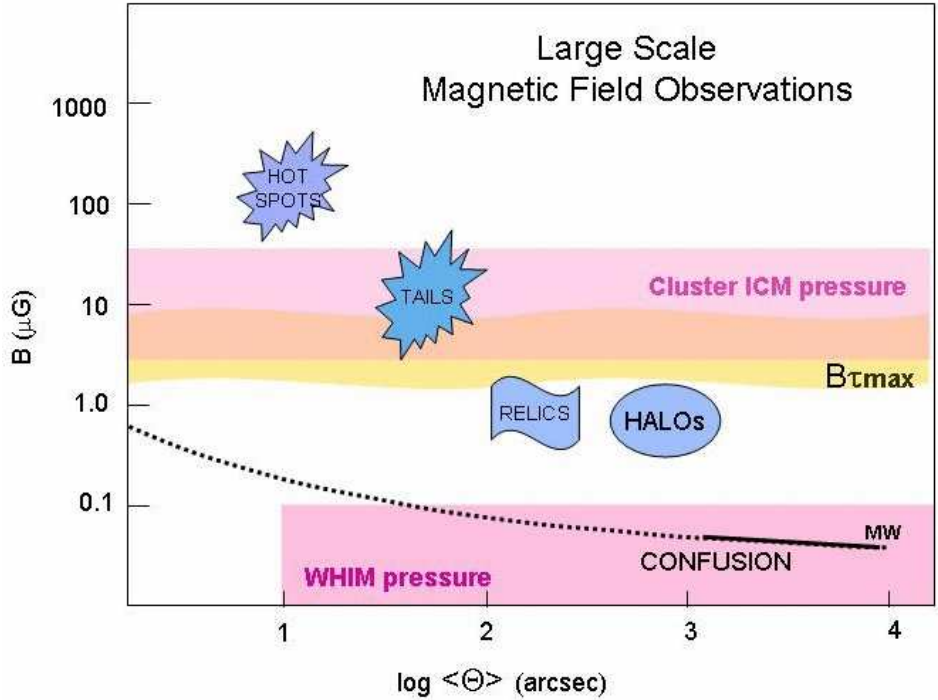}}
\vskip -0.2cm \caption{[B,$\theta$] space - a perspective on
observed structures, pressure-equivalent benchmarks, and 
observational constraints.  The curved dashed line towards the
bottom represents the confusion limit, irreducible in the
large angle (solid line) portion due to the Milky Way.}
\vskip -0.5cm
\label{fig1}
\end{figure*}

Two "benchmark" field strengths/pressures  are indicated -- those associated
with the thermal pressure in clusters of galaxies, which range over
an order of magnitude across a cluster (e.g. Briel, Fignogeunov \& Henry, 2004)
 and pressures characteristic of the WHIM (e.g. Nicastro 2003).  Eventually, 
when the scaling relationships are better understood for these
pressure benchmarks, it would be useful
to show their $\theta$ dependence (perhaps at different redshifts) on the [B, $\theta$] diagram.

This now brings us to the question of observational constraints -- the main
reason for constructing the [B, $\theta$] diagram.   The first constraint
is provided by the \emph{maximum lifetime field} (B$_{\tau max}$)- the
field strength which maximizes the radiative lifetime of emission
\emph{at a fixed observing frequency}.  Ignoring Coulomb losses,
which are unimportant in the radio regime for diffuse synchrotron
sources, and assuming an initial power law spectrum of $-0.5$,
B$_{\tau max} = 3.2~(7.1)~\mu$G at $z = 0~(0.5)$, corresponding to
lifetimes of $10^{8.3~(8.0)}$ years.  At field strengths a factor of
$\approx$4 from B$_{\tau max}$, the lifetimes drop by a factor of two.
A maximum lifetime field exists because if the field were
higher, synchrotron losses would reduce the lifetime, and if the
field were lower, synchrotron emission at a given observational
frequency would come from higher energy particles, whose inverse
Compton losses would again reduce the lifetime. Curiously, the
maximum lifetime \emph {field strength} is independent of observing frequency (although
the lifetime itself is frequency dependent).

The \emph{maximum lifetime field} is closely related to the
\emph{maximum lifetime particle energy} (E$_{\tau max}$), as discussed
over the years and described in the tutorial by Sarazin (1988).  The
distinction between the two is important: B$_{\tau max}$ tells you what
strength fields containing a fixed population of particles 
will be observed for the longest times;  it thus
forces an observational selection bias as shown in Figure 1.
 E$_{\tau max}$ tells you the
energy and lifetime of the longest lived relativistic particles, which 
 can later be
re-energized to become visible at radio frequencies.  This visibility
depends on the re-energization and the  local current magnetic field strength, and provides a
different set of model-dependent selection effects. 
This paragraph is worth re-reading.

The second critical observational constraint is related to
 ``confusion'' levels (Condon, 1974). To study very low
surface brightness features, one typically wants to work
at low resolution, approximately on the angular scale of the
target source.  If the observations
are sensitive enough, the fluctuations in the background
due to the combined flux of sources within one ``beam''
then become dominant.  In Figure 1, I have indicated this
confusion limit in terms of its equivalent magnetic field
strength (B$_{min}$), assuming an equal contribution from relativistic
particles  at each angular scale (as opposed to physical scale).
  This is a useful approximation, because the redshift
dependence of B$_{min}$ is only $z^{2/7}$ at a fixed confusion
flux limit and a fixed angular scale.

  Unfortunately, this confusion limit is right in the
regime where relativistic plasmas in pressure equilibrium with the
WHIM would become visible.  Attempts to push below the confusion
are described below, although at sufficiently large angular scales
$\ge1^o$, the irreducible confusion from the Milky Way will 
become dominant.

\section {Rotation Mis-measures}

Since there are widespread  magnetic fields and thermal plasmas in  clusters of
galaxies,  at some level their resulting Faraday rotation will
be seen.  However, the evidence to date that observed rotation
measures are due to cluster wide fields (with magnitudes of
 1-35$\mu$G, Carilli \& Taylor 2002) is on quite shaky
grounds. 

\begin {figure*}[t]
\vskip 0cm \centerline{\epsfysize=7.cm\epsfbox{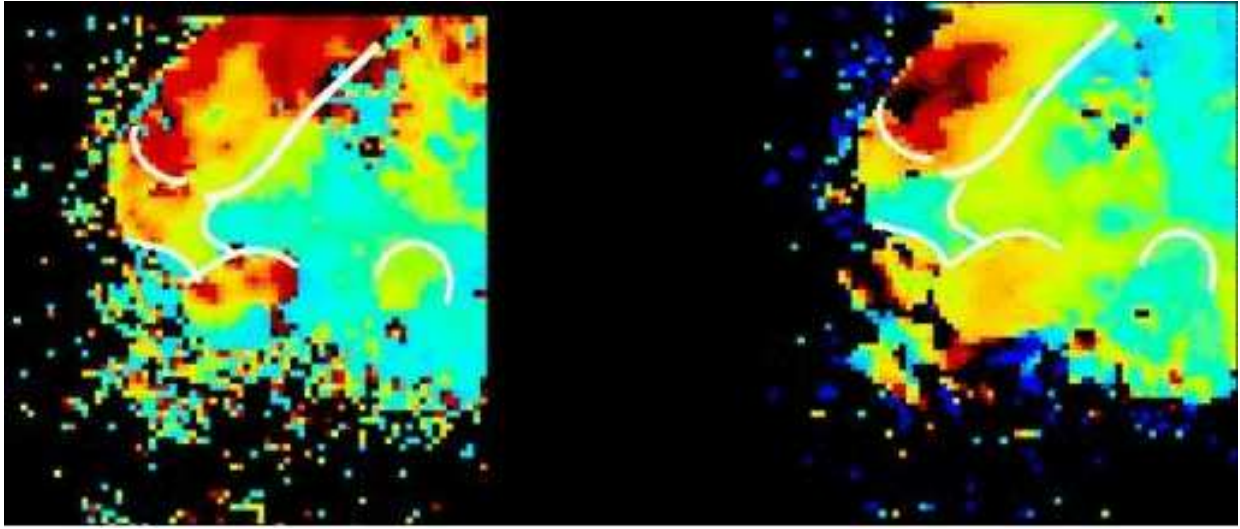}}
\vskip -0.2cm  \caption{Polarization of the eastern lobe of Cygnus A, original and derived maps
 kindly provided by
C. Carilli \& R. Perley -- these maps were derived from multifrequency data
at 3.7cm  using the VLA.  Left: Magnetic field orientation, color coded; Right: Rotation Measures (arbitrary color scale). Hand-drawn white lines indicate positions where rapid shifts in
magnetic field direction are accompanied by rapid shifts in Rotation Measure -- an unambiguous signature of rotation measure \emph{ local} to the source, not due to any intervening cluster field.}
\vskip -0.5cm
\label{fig2}
\end{figure*}

There are two types of cluster rotation measure studies.  In the
first, rotation measure variations across an individual cluster
radio source are used to infer cluster-wide fields. 
Rudnick \& Blundell (2003) discuss the numerous problems with
this inference, and argue that a plausible alternative is that
the rotation measures arise in a thin thermal skin mixed with the
radio source's own
relativistic plasma.  An excellent example of this effect is shown
in Figure 2.  Here we see the polarization angle structure of
the eastern lobe of Cygnus A (which must be intrinsic to the source)
and the corresponding rotation measure structure (see Perley \& Carilli 1996).  It is obvious
that many patches of rotation measure correspond to coherent
patches of Cygnus A's own magnetic field -- \emph{ not} some random
foreground cluster screen.  Until such effects are eliminated
from maps of individual sources (and there is no way presently
to do so), one cannot infer cluster-wide fields from such observations.
The demonstration that certain cluster-wide field geometries \emph{ might}
explain variations across \emph{ some} sources (Ensslin et al. 2003) falls quite far short
of demonstrating that cluster fields are actually responsible for observed
rotation measures.  

The second line of evidence claimed for cluster-wide fields is
the larger dispersion of rotation measures for distant background
sources seen through clusters than for control samples not seen
through clusters.  While in principle this method could work,
the studies to date are seriously flawed (see Rudnick \& Blundell, 2004).

An example of a flawed statistical study  is shown in Figure 3.  The claims for an excess rotation
measure in the direction of the Coma Cluster (Kim et al. 1990) are
based on what appears to be a statistically significant excess near
the cluster center.  Unfortunately, the sample contains many sources
that are actually in the cluster itself (a demonstrably dense
environment with distinct types of radio sources) as opposed to the
control sample of background radio sources. This flaw in the
experimental design must be corrected by removing actual cluster
sources from the sample.  In addition, there are multiple serious
apparent errors in the actual data analysis.  5C4.70, for example,
with an 18cm polarization percentage of 0.9$\pm$0.7 (a non-detection)
leads to only a 5$^o$ error in polarization angle (Kim et al. 1994),
and then to a highly accurate quoted rotation measure in Kim et
al. (1990).  Similarly, 5C4.112b has a reported 21cm polarization
percentage of 0.9$\pm$1.0 (a non-detection) a polarization angle error
of 10$^o$ and another high quality quoted rotation measure.  The
results of eliminating cluster sources and re-calculating errors is
shown in Figure 3 for the Kim et al. work.  It is obvious that the
Coma cluster magnetic field, known for decades through its synchrotron halo
(e.g., Willson 1970), shows no measureable rotation measure effect at
present levels of sensitivity.

\begin {figure}[t]
\vskip 0cm \centerline{\epsfysize=6.cm\epsfbox{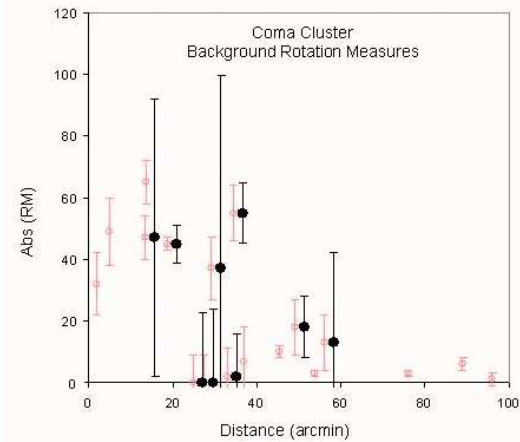}}
\vskip -0.2cm  \caption{Rotation Measures of radio sources
in the direction of the Coma Cluster from Kim et. al 1990 (red). The
same data (offset for clarity), but including only actual background sources, and showing the correct errors (black). No cluster RM effect is visible in the corrected data.}
\vskip -0.5cm
\label{fig3}
\end{figure}

At this meeting, T. Clarke and M. Johnston-Hollitt each defended the
statistical studies.  Clarke presented revised data from Clarke,
B\"{o}hringer \& Kronberg (2001), by appropriately dropping the
sources actually embedded in clusters.  However, to improve the now
poor statistics, she added data from the literature 
of questionable quality (e.g., unmatched
spatial resolution, errors in data analysis) such as noted above
from Kim et al. (1990).  Johnston-Hollitt showed the results from a
new southern hemisphere survey; her analysis is necessarily
complicated since the sources are spatially
resolved. However, there is no control sample, no off-cluster sources
that are observed and processed in the same way, so it is not possible
to draw any conclusions about clusters from this work.
 
There  thus remain no reliable measurements
of cluster-wide fields from rotation measure studies at present.
With a large investment of time, it might be possible to do this
background source experiment properly using the Very Large Array.  Otherwise, the SKA offers the most promise.

\section{Relics}
Studies of these patchy cluster sources without an obvious parent AGN have
reached the point that they can be classified into physically meaningful
categories (see Kempner et al. 2003). They distinguish between:\\
 a) ``AGN Relics'' ($<$100 kpc) which may actually show connections to
 a nearby  AGN when deep maps are available, and are
likely associated with bouyant bubbles and dredged up thermal material from the
parent cD core, and\\
b) ``Radio Phoenix'' structures on larger scales,  which are presumed to illuminate relativistic
plasma from previous radio galaxy activity, currently 
recompressed and reaccelerated by merger or other shocks; and \\
c) ``Radio Gischt'' - thin relics up to several Mpc long on the
periphery of clusters, resulting from either merger or accretion shocks.\\
In these proceedings, a nice overview of relic properties, with a
somewhat different classification scheme, is presented by G. Giovannini.

In this paper, I present initial spectral studies of one AGN Relic and
one Radio Phoenix from the thesis work of A. Young. The purpose of these
studies is to uncover the  acceleration processes
that provide the initial population of relativistic electrons.

\begin{figure}[t]
\vskip 0cm \centerline{\epsfysize=6.cm\epsfbox{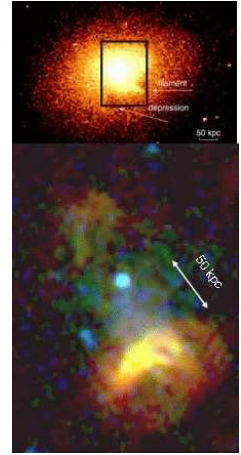}}
\vskip -0.2cm \caption{Top: Chandra (XMM) image of the
inner 200 kpc of the MKW3s cluster, centered on the cD (from Mazotta et al. 2002). The box indicates
the approximate region shown in radio emission at the bottom, from multifrequency VLA maps) color coded
by spectral index, with red=steep (Young et al. 2004).}
\vskip -0.5cm
\label{fig4}
\end{figure}

\subsection{An AGN Relic}

The MKW3s system is shown in Figure 4.  The X-ray emission shows a
bright finger of emission emerging to the Southwest, extending out
over 50 kpc from the core.  The temperature of this plasma is
considerably lower (by 2-3 keV) than the surroundings, and is fairly
well aligned with the radio structure.  The bright steep-spectrum
filamentary emission at the Southwest terminus of the radio source
was seen in the FIRST survey (Becker, White \& Helfand 1995), but its
origin was unclear (Mazzotta et al. 2002).  These new deep observations,
at 90, 20, and 6cm on the VLA\footnote{The Very Large Array is a facility of the U.S. National Science Foundation, operated by the National Radio Astronomy Observatory under contract through Associated Universities, Inc.} find a bridge of emission leading back through an AGN core and extending to another faint
steep-spectrum feature at the northern terminus.

\begin{figure*}[t]
\vskip 0cm \centerline{\epsfysize=6cm\epsfbox{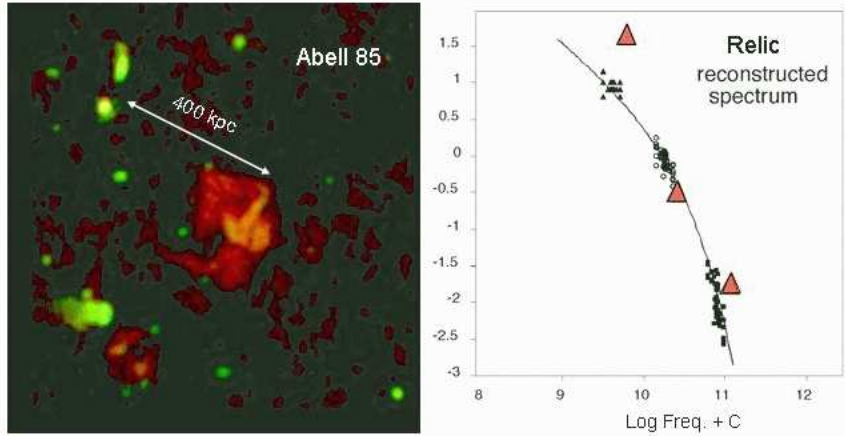}}
\vskip -0.2cm \caption{Left: VLA radio image of Abell 85 (Young et al. 2004) at a resolution
of 14'', with color-coded spectral indices (red=steep).  Right: Reconstructed
spectrum of the relic as described in the text, along with triangles showing
the  relative sensitivities at different VLA observing frequencies.}
\label{fig5}
\end{figure*}

The original filamentary radio source, which previously could have
 been thought of as unpowered ``relic'' emission, 
 has therefore been transformed into a radio galaxy
type structure, albeit a most curious one, with extremely steep
( $<-$3.5!) spectra at its bright end.  Physically,
it must have a different origin than the ``relaxed doubles'' or the
tailed cluster radio galaxies that fade and steepen away from the nucleus. It
must also have a different origin than jet driven high luminosity doubles
that have bright flat spectrum regions at their terminus.  Nor is it
simply a bouyantly rising detached bubble, since there is an AGN and
a continuous bridge of emission.  

This radio/X-ray system is very similar to that of Abell 133 (Fujita
et al. 2002), although the structure of the diffuse emission is not 
yet clear.  These two 
systems both have steep spectrum filamentary radio structures at the
end of a cool column of thermal gas, extending $\approx 50 kpc$  from the core of a cooling flow cluster.

One possibility for these unusual physical systems is that they represent
a  continuous supply of energy from the AGN, but with a wider opening angle
than the collimated jets often seen in radio galaxies.  Such a broad flow,
probably subsonic, could entrain and dredge up the cool bright finger
of thermal gas, as well as accumulate a large, bright  pool of relativistic
particles at the terminus, strongly steepened by radiative losses.
Simulations of continuously inflated MHD  bubbles (Jones \& De Young 2004)
 show that they
can develop mildly supersonic flows and filamentary 
magnetic field structures, and may be a promising way to explain sources such
as MKW3s and Abell 133. 

\subsection{A Radio Phoenix}

Abell 85 is a well-studied dynamic cluster, with multiple on-going
mergers (Kempner, Sarazin \& Ricker 2002; Durrett et al. 2003).  The
steep-spectrum radio relic (the prototype ``Phoenix'', Kempner et
al. 2003) has been mapped by Bagchi, Pislar \& Lima Neto (1998) and
Slee et al. (2001).  A wide field view of the cluster, showing the
steep spectrum relic and flatter spectrum radio galaxies both in and
projected onto the cluster is seen in Figure 5.  Our multifrequency
VLA observations were designed to determine the spectral shape. on a
beam-by-beam basis.  Although obtaining integrated spectra are much
easier, and a good first step (e.g. Bagchi et al. 1998), the
beam-by-beam spectra are a truer representation of the actual electron
distributions because they avoid spectral smearing.  Our reconstructed
spectrum is shown in Figure 5. It is constructed by sliding the
observations for each beam by arbitrary amounts in ($logI~vs.~log\nu$)
to fit onto the spectral shape derived from a color-color analysis
(Katz-Stone, Rudnick, \& Anderson 1993). It has an extrapolated \emph{
low-frequency index} of $-$0.87, much steeper than any other radio
galaxies.  One interpretation is that the relativistic electrons did
not originate in a radio galaxy, but have been newly accelerated in a
low Mach number shock.  The spectral shape data may also be compatible
with the flatter low frequency spectra of  radio
galaxies, if there are significant fluctuations in magnetic field along
each line of sight.  These possibilities are explored further in Young
et al. (2004).

A very important observational lesson is also apparent from Figure 5.  The
triangles denote the relative sensitivities of the VLA at wavelengths of
90, 20, and 6 cm. If you consider the observed spectrum as a model
source, you can slide it to any arbitrary position in $log I~vs.~ log
\nu$ space to determine its detectability.  The important thing to
note is that for a source with this amount of curvature, emission may
easily be detectable at 20cm, but not at either 90cm or 6cm.  Thus,
one cannot rely on low frequency surveys, for example, to detect such
sources, even though they have very steep spectra.  Curvature in the
spectrum and varying sensitivities demand that searches be conducted at
a variety of frequencies.

\section{Unbiased searches for diffuse emission}

\subsection{WENSS/WISH}

Since previous searches for halos and relics have focussed primarily
on clusters (e.g. Kempner \& Sarazin 2001, Giovannini \& Feretti
2002), we have undertaken an unbiased search of the WENSS (Rengelink
et al. 1997) and WISH (the southern extension, De Breuck et al. 2002)
surveys for diffuse sources (Delain et al. in preparation).  The basic
technique was to filter all the images from these surveys to remove
the small scale structure and search the residuals, using the simple
multiresolution filtering scheme described by Rudnick (2002).  A wide
variety of diffuse sources were found, many of them already known.
These included giant radio galaxies and cluster halos and relics. 

One of the newly recognized sources is shown in Figure 6 (without
coordinates as we attempt to understand the optical environment).  The
radio structure consists of two patches of diffuse emission that
extend roughly perpendicular to each other over more than 10'.  There
is no extended X-ray emission visible, although the SDSS shows the
presence of several loose clusters or galaxy groups.  This system
suggests that diffuse radio sources can form outside of dense
potential wells, and therefore raises questions about the mechanisms
currently invoked for producing them. It emphasizes that searches for
diffuse emission need to look beyond rich and/or X-ray emitting
clusters. 

\begin {figure}[t]
\vskip 0cm \centerline{\epsfysize=6.cm\epsfbox{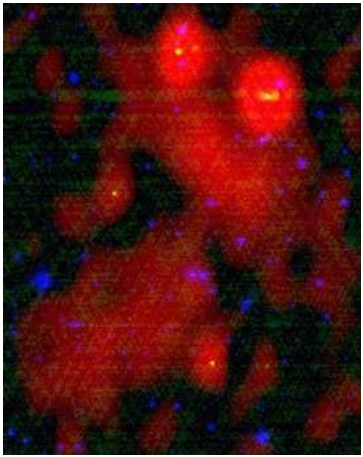}}
\vskip -0.2cm  \caption{A diffuse radio source discovered in the
WENSS survey (red) overlaid with the FIRST images of compact sources
(yellow/green) and the Digital Sky Survey (blue).}
\vskip -0.5cm
\label{fig6}
\end{figure}

\subsection{TONS08}

\begin {figure*}[t]
\vskip 0cm \centerline{\epsfysize=6.cm\epsfbox{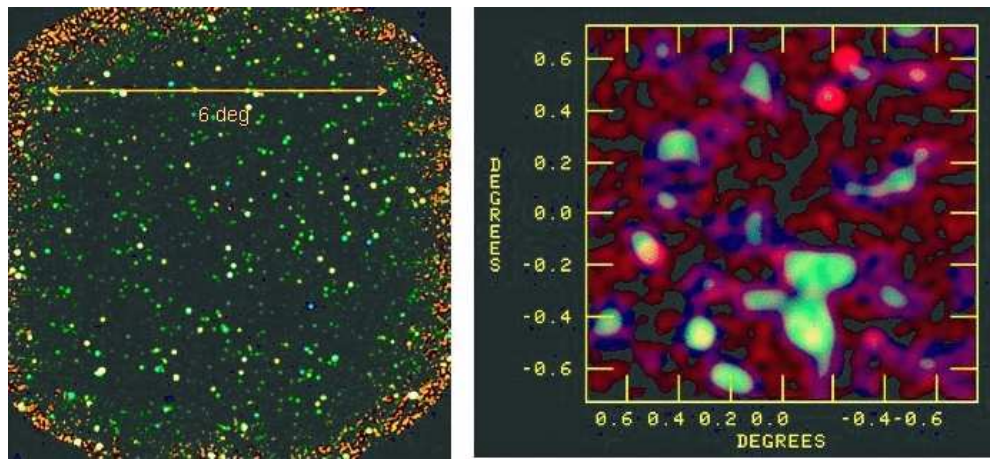}}
\vskip -0.2cm  \caption{The TONS08 wide field VLA mapping at
  90cm. Left: full 6$^o\times$6$^o$ field color coded by spectral
  index using the NVSS (red=steep).  Right: closeup of residual
  emission showing possible diffuse structure in green.}
\vskip -0.5cm
\label{fig7}
\end{figure*}
We are conducting another blind search for diffuse radio structures by
making deep VLA maps at 90cm of the TONS08 area from the TexOx-1000
(TOOT) Project \footnote
{http://www-astro.physics.ox.ac.uk/~sr/texox/toot.html}. This field is
being studied by others in a variety of  ways, and two 100 Mpc
superstructures have already been identified (Brand et al. 2003).  The
initial results from the new radio work are shown in Figure 7.  On the
left is a 36 square degree (6$^o\times$6$^o$) field constructed as a
mosaic of 26 individual telescope pointings.  The resolution
of this image, made using the VLA D configuration, is $\approx$180'';
it is thus sensitive to sources much larger than can be seen in the
NVSS (Condon et al. 1998).  The color provides a
rough measure of the spectral index (red=steep) by comparison with the
NVSS survey  of the same region. 

To maximize the sensitivity to large scale structures we need to
convolve this image to a lower resolution.  This immediately takes us
to the ``confusion'' limit discussed above and shown in Figure 1.  In
order to reach below the confusion, we therefore conducted the same
set of observations using the VLA B configuration, at approximately
10$\times$ finer resolution, and subtracted the contributions from
small sources from the low resolution data.  

The result of convolving these residuals to a resolution of 420'' is
shown on the right of Figure 7. Approximately 90\% of the flux from
small sources has been removed, and we are working to improve this
subtraction.  Through the magic of image processing, a 1.5$\sigma$
diffuse source is visible in green at the position (-0.2,-0.2).  The
convolved original map is shown in red; yellow regions indicate the
presence of both compact sources and diffuse residuals.  Given the
variety of effects that produce sidelobes and other spurious
structures in deep radio images, investigations such as these must be
done quite carefully.  However, our results so far are encouraging and
we look forward to probing the lower pressure regions associated with
large scale structures.

\section{Concluding Remarks}
In this paper, I have shown the limited view we have today of magnetic fields on large scales, and some of
the opportunities for widening our knowledge, both with current
instruments and especially with the new low frequency arrays and the
SKA.  It appears likely that  magnetic fields can grow and
relativistic particles can be accelerated under a wider range of
physical conditions than currently observed.  As we probe these new
regimes, it is incumbent upon us to work rigorously to ensure that our
observations reflect the universe, and not simply our expectations.

\acknowledgements{This work was supported in part by the U.S. National
Science Foundation through grant AST-0307600 to the University of
Minnesota.  It is a pleasure to acknowledge the work of many people
that have led to the results shown here.  The mapping and spectral
work on radio relics (MKW3s and Abell 85) were conducted by Andrew
Young as part of his thesis work, and are being studied further in
collaboration with H. Andernach (Univ. Guanajuato, Mexico), N. Kassim
(NRL), J. Kempner (Bowdoin), A. Roy (MPIfR, Bonn), C. Sarazin
(Virginia), and B. Slee (ATNF).  Rotation measure critiques were done
in collaboration with K. Blundell (Oxford).  The technically
challenging work on searches for diffuse radio emission is being done
by graduate student Kisha Delain (U. MN). The WISH survey was kindly
provided in electronic form by C. de Breuck (Institut d'Astrophysique
de Paris).  Studies of the TONS08 region are done in collaboration with
S. Rawlings, K. Brand, and K. Blundell (Oxford). Thanks to C. Carilli
\& R. Perley for use of unpublished versions of their 3.7cm
polarization images of Cygnus A.  Tom Jones provides expertise,
encouragement and constructive criticism, all as appropriate.
Finally, any unwarranted speculations or impolitic remarks are the
responsibility of the author alone. }

\end{document}